\documentclass[singlecolumn,showpacs,preprintnumbers,amsmath,amssymb,aps]{revtex4}
\usepackage{graphicx}
\usepackage{dcolumn}
\usepackage{bm}
\begin{document}
\preprint{}
\title{Photon emission from out of equilibrium dissipative parton plasma}
\author{Jitesh R. Bhatt{\footnote {email: jeet@prl.res.in}} and V. Sreekanth{\footnote {email: skv@prl.res.in}}}
\affiliation { Theoretical Physics Division, Physical Research Laboratory, Navrangpura, Ahmedabad, India - 380 009}
\date{\today}
\begin{abstract}

Using the second order Israel-Stewart hydrodynamics we discuss 
the effect of viscosity on photon production in a parton
plasma created in relativistic heavy ion collisions.
We find that photon production rates can enhance by several
factors due to the viscous effect in a chemically nonequilibrated
plasma.

\end{abstract}
%\pacs{97.60.Jd, 04.30.Db, 04.40.Dg, 26.60. +c}
\maketitle
%\section{INTRODUCTION}

A strongly coupled quark-gluon plasma (sGGP) or a matter in a
perfect fluid state is widely expected to be produced in recent
Relativistic Heavy Ion Collider (RHIC) experiments. Recent measurements has shown that the matter flows very rapidly at
the time of its breakup into the freely streaming hadronic matter.
Also the measurements 
of the elliptical flow parameter $v_2$ show a strong collectivity in
the flow \cite{star,phen,phobos}. 
This would imply that the QGP can have a very low shear
viscous stress and  the ratio of its shear viscosity $\eta$ 
to the entropy density $s$ i.e. $\eta/s$ should not be
much larger than  the lower bound $1/4\pi$ 
 \cite{kss05}. This led to a conjecture that the QGP
formed at RHIC is the most perfect-fluid found
in nature \cite{Hirano:2005wx}. There has been a lot of attempts
to determine the viscosity of sQGP \cite{Hirano:2005wx, bmpr90,
hk85, heisel94, AYM, XUPRL}.
 The first order theory of viscous hydrodynamics is known to
give unphysical results. For example, when the Navier-Stokes
equations were applied to the one dimensional boost invariant
expanding flow \cite{bjor}, one finds the expression for the temperature
to be 
$$
T(\tau)=T_0 \left(\frac{\tau_0}{\tau}\right)^{1/3}\left[1+\frac{2\eta}{3 s
 \tau_0 T_0}\left(1-\left(\frac{\tau_0}{\tau}\right)^{2/3}\right)\right].
 \nonumber
$$
This describes  reheating of the flow  and  $T$
has a maximum at time $\tau_{\rm max} = \tau_0 \left( \frac{1}{3} + \frac{s}{\eta} \frac{\tau_0\, T_0}{2} \right)^{-3/2} $.
This is an unphysical behaviour and the first order viscous
dynamics is known to have such problems 
\cite{lindblom,Baier:2006um}. The second order hydrodynamics 
approach developed in the spirit of Israel and Stewarts \cite{Israel:1979wp} removes such  artifacts.
The second order viscous hydrodynamics later developed
and applied in the context of heavy-ions collisions
by several authors
\cite{br07,Teaney03, SongHeinzCh06,hs08,Hama01,AM207,R97,DT08}.

It would be interesting to study the role that viscosity
can play on the plasma signals.
Hard photons are one such promising source
that can provide information about the thermodynamical
state of the  plasma at time of their production.
The plasma created in the heavy-ion collisions is expected to
be in a state of chemical nonequilibrium. The photon emission
from such a plasma has been studied within the framework of ideal
hydrodynamics by earlier workers \cite{thoma,dmkc00,bdrs1997,long05}.
In this paper we study the photon production using 
causal hydrodynamics of Israel-Stewart \cite{Israel:1979wp}.

In the center of the fireball in a nuclear collision the viscous 
stress-energy tensor in the local comoving frame has the form 
 \cite{Teaney03, Muronga:2001zk,Muronga:2004sf}:
\begin{equation}
T^{\mu\nu} = \left(
\begin{array}{cccc}
\varepsilon & 0 & 0 & 0 \\
0 & P_{\perp} & 0 & 0 \\
0 & 0 & P_{\perp} & 0 \\
0 & 0 & 0 & P_{z} 
\end{array} \right) 
\label{eq:Tmunu}
\end{equation}
with the transverse and longitudinal pressure
\begin{eqnarray}
P_{\perp} &=& P + \frac{1}{2}\Phi 
\nonumber \\
P_{z} &=& P - \Phi
\label{eq:stress}
\end{eqnarray}
Here $P$ denotes the (isotropic) pressure in thermal equilibrium, 
$\Phi$ denotes the non-equilibrium contributions to the pressure
coming from shear stress. We ignore the bulk viscosity
in the relativistic limit
when the equation of state $p=\epsilon/3$ is obeyed \cite{wein}.
However, the bulk viscosity can be important near the critical
temperature \cite{fms08, karsch07}.
The shear tensor in that frame takes the form 
$\pi^{ij} = \mathrm{diag}(\Phi/2, \Phi/2,-\Phi)$
consistent with the symmetries in the transverse directions.

To describe evolution of the energy density and the viscous stress $\phi$
we use second order dissipative hydrodynamics of Israel-Stewart
\cite{Israel:1979wp,Heinz:2005zi,Muronga:2003ta,emxg08}:
\begin{eqnarray}
  \frac{\partial\varepsilon}{\partial\tau} &=& - \frac{1}{\tau}(\varepsilon 
  + P  - \Phi) \, ,  
  \label{eq:evol1} \\
  \frac{\partial\Phi}{\partial\tau} &=&  -\frac{\Phi}{\tau_{\pi}}-\frac{1}{2}\Phi \left( \frac{1}{\tau} + \frac{1}{\beta_2}T\frac{\partial}{\partial\tau}(\frac{\beta_2}{T}) \right) + \frac{2}{3}\frac{1}{\beta_2\tau} \,,
\end{eqnarray}  
where $\beta_2=9/(4\varepsilon)$ and $\tau_{\pi}=2\beta_2\eta$ denotes 
the relaxation time.
Equations(3-4) are written in the local rest frame using hydrodynamic
velocity $u^\mu=\frac{1}{\tau}(t,0,0,z)$, where 
$\tau=\sqrt{t^2-z^2}$ \cite{bjor}. Equation of state is required to solve
these equations. We use ultra-relativistic equation of state
: $P=\frac{1}{3}\epsilon$.

To describe the chemical non-equilibration while maintaining the
kinetic equilibrium, one can use the parton distribution \cite{levai},

\begin{equation}
f(k,T)_{q,g}\,=\, \lambda_{q,g}(\tau)\frac{1}{e^{{\mathbf {u\cdot k}}/T(\tau)}\,\pm\,1}
\end{equation}

\noindent 
where, $u^\mu$ is the four-velocity of the local comoving reference frame.
The temperature $T$ is a time-dependent quantity and the distribution is multiplied by
time and another dependent quantity called fugacity $\lambda_{q,g}(\tau)$ to describe deviations
from the chemical equilibrium. The fugacity parameter become unity when the chemical-equilibrium is 
reached and in general it has the range $ 0\le \lambda_{q,g}\le 1$. The scattering processes
$gg\leftrightarrow ggg $ and $gg\leftrightarrow q\bar{q}$ give the most dominant mechanism
for the chemical equilibration. The master equations describing evolution the parton density are
given by
\begin{eqnarray}
\partial_\mu(n_gu^\mu)&=&\frac{1}{2}\sigma_3n^2_g\left(1-\frac{n_g}{ \tilde{n}_g}\right)
-\frac{1}{2}\sigma_2n^2_g\left(1-\frac{n^2_q\tilde{n}^2_g}{\tilde{n}^2_qn^2_g}\right),\\
\partial_\mu(n_qu^\mu)&=&\frac{1}{2}\sigma_2n^2_g
\left(1-\frac{n^2_q\tilde{n}^2_g}{\tilde{n}^2_qn^2_g}\right),
\end{eqnarray}
\noindent where $\tilde{n}_i$($i=q,g$) is parton density with unit fugacity \cite{biro}
and $\sigma_2\,=\langle\sigma(gg\leftrightarrow q\bar{q})\rangle$ and
$\sigma_3\langle\sigma(gg\leftrightarrow ggg)\rangle$ are thermally averaged 
scattering cross sections.  It should be noted here that
when equation for $n_g$ and $n_q$ are added one gets the total
number density $n$ and the term with $\frac{1}{2}\sigma_2n^2_g
\left(1-\frac{n^2_q\tilde{n}^2_g}{\tilde{n}^2_qn^2_g}\right)$
will drop out. This is because due to the the scattering process
$gg\leftrightarrow q\bar{q}$ loss in the gluon density is equal to the gain
in quark density and vice verse. 

\noindent
$\epsilon$ and  $n$  can be calculated using equation (5) as given below
\begin{eqnarray}
n=(\lambda_g a_1 + \lambda_qb_1)T^3 , \,\,\,\,\epsilon=(\lambda_g a_2 + \lambda_q b_2)T^4
\end{eqnarray}
\noindent
where $a_1={16\xi(3)}/{\pi^2}$, $a_2=8\pi^2/15$ for the gluons and $b_1=9\xi(3)N_f/\pi^2$,
$b_2=7\pi^2N_f/20$ for the quarks. Using equations (3-4,6-8) following
evolution equations for $T$, $\lambda_{q,g}$ and $\Phi$ can be obtained
\begin{widetext}
\begin{eqnarray}
\frac{\dot{T}}{T}+\frac{1}{3\tau}&=&-\frac{1}{4}
\frac{\dot{\lambda}_g+b_2/a_2\dot{\lambda}_q}
{{\lambda}_g+b_2/a_2{\lambda}_q} + \frac{\Phi}{4\tau}\frac{1}{(a_2{\lambda}_g+b_2{\lambda}_q)T^4}
  \, ,  
  \label{eq:tempevol1} \\
\dot{\Phi}+\frac{\Phi}{\tau_\pi}&=&\frac{8}{27\tau}
\left[ a_2\lambda_g+b_2\lambda_q\right]T^4-\frac{\Phi}{2}
\left[\frac{1}{\tau}-5\frac{\dot{T}}{T}-
\frac{\dot{\lambda}_g+b_2/a_2\dot{\lambda}_q}
{{\lambda}_g+b_2/a_2{\lambda}_q}\right]
\, ,\label{eq:shearvisco4}\\
\frac{\dot{\lambda_g}}{\lambda_g}+3\frac{\dot{T}}{T}+\frac{1}{\tau}
&=&R_3\left(1-\lambda_g\right)-R_2\left(1-\frac{\lambda^2_q}
{\lambda^2_g}\right)
\, ,
    \label{eq:gluonfugacityevol2}\\
\frac{\dot{\lambda_q}}{\lambda_q}+3\frac{\dot{T}}{T}+\frac{1}{\tau}
&=&R_2\frac{a_1}{b_1}
\left( \frac{\lambda_g}{\lambda_q}-\frac{\lambda_q}{\lambda_g}
\right)
    \label{eq:quarkfugacityevol3} 
\end{eqnarray}
\end{widetext}
where,  the rates 
$R_2=0.24N_f\alpha^2_s\lambda_gTln(5.5/\lambda_g)$
and $R_3=2.1\alpha^2_sT(2\lambda_g-\lambda^2_g)^{1/2}$
are defined as in Ref. \cite{levai,biro}. We would like to note that our gluon fugacity equation (\ref{eq:gluonfugacityevol2}) differs from that given in Ref. \cite{levai,biro} by a fctor of two in second term in right hand side. We believe this is a typographical error. In equation (\ref{eq:tempevol1}) the first term on left hand side is due to expansion of the plasma, while  on the right hand side the first term describes effect of chemical nonequilibrium and second term is due to the presence of (causal) viscosity. The last term in parenthesis of equation (\ref{eq:shearvisco4}) arises because of 
the chemical nonequilibrium process. 
It should be noted that equation (9)
differ from that considered in Ref. \cite{chau00}. In their treatment
first order viscous hydrodynamics is used which does not require
time evolution of $\Phi$. However such treatment give unphysical
results like reheating artifact \cite{Baier:2006um} as mentioned before. 

Elastic $(gg\leftrightarrow gg)$  as well as nonelastic
processes like $gg\leftrightarrow ggg$ can contribute to the shear viscosity.
Shear viscosity coefficient was recently calculated for the inelastic process
in the presence of chemical nonequilibrium in Ref. \cite{emxg08}. It was shown
that $\eta/T^3\simeq\,n_g/T^3\simeq \lambda_g$. From this one can write \cite{Muronga:2003ta}
 \begin{equation}
 \tau_{\pi}=\frac{9}{2\varepsilon}\lambda_gT^3 .\label{taupimronga}
\end{equation}
It ought to be mentioned that this viscosity prescription
was not considered considered in Ref.\cite{chau00}.
Kinetic theory without invoking nonequilibrium process
gives $\tau_{\pi}=3/{2\pi T}$.

Real photons are produced from the annihilation of a quark-antiquark pair
into a photon and a gluon ($q\bar q\rightarrow g\gamma$)
and by absorption of a gluon by a quark emitting
a photon ($qg\rightarrow q\gamma$). Another source of photon production
could be the bremsstrahlung but its effect can be ignored in the lowest
order of a perturbation theory.
In order to compute the photon production rates one needs to know
the underlying amplitude $\mathcal{M}$ of the basic process involving the annihilation
or Compton scattering process and the parton distribution functions given by \cite{peitzmann,gh03}
\begin{eqnarray}
%\hspace*{-1cm}
\frac{dN}{d^4xd^3p}&=&\frac{1}{(2\pi)^32E}\> 
\int \frac{d^3p_1}{(2\pi)^32E_1} 
\frac{d^3p_2}{(2\pi)^32E_2} \frac{d^3p_3}{(2\pi)^32E_3}\> \nonumber\\
&&
 \times n_1(E_1)n_2(E_2)[1\pm n_3(E_3)] \\ \nonumber
&& \times \sum_{i} \langle |\mathcal{M}|^2\rangle \>
(2\pi)^4\> \delta(P_1+P_2-P_3-P).
\label{rate1}
\end{eqnarray}
Here $P_1$ and $P_2$ are the 4-momenta of the incoming partons, 
$P_3$ of the outgoing parton, and $P$ of the produced photon.
In equilibrium, the distribution functions 
$n_i(E_i)$ are given by the Bose-Einstein distribution, 
$n_B(E_i)=1/[\exp(E_i/T)-1]$, for gluons and by the Fermi-Dirac distribution,
$n_F(E_i)=1/[\exp(E_i/T)+1]$, for quarks, respectively.
The factor $\langle |\mathcal{M}|^2\rangle$
is the matrix element of the basic process averaged over the initial states
and summed over the final states. The $\sum_{i}$ indicates the sum over
the initial parton states.
The fugacity factors can enter equation (14) when equation (5) is considered\\
\begin{equation*}
n_1(E)n_2(E)(1\pm\,n_3(e))\mapsto\,
\lambda_1n_1\lambda_2n_2(1\pm\,\lambda_3n_3) \nonumber.
\end{equation*}
This is can be rewritten as
\begin{eqnarray}
 \lambda_1n_1\lambda_2n_2(1\pm\,\lambda_3n_3)&=&
\lambda_1\lambda_2\lambda_3
n_1n_2(1\pm\,n_3)\\ \nonumber
&+&\lambda_1\lambda_2(1-\lambda_3)n_1n_2
\end{eqnarray}
\noindent
In carrying out the momentum integration it is useful to introduce
a parameter $k_c$ to distinguish between soft and hard momenta of the
quark \cite{kapu91}. For the hard part of the photon rate 
following \cite{thoma} we take 
$k^2_c=2m^2_q=0.22g^2T^2\left(\lambda_g+\lambda_q/2\right)$,
where, $m_q$ is the quark-thermal-mass which can be obtained from
zero momentum limit of quark self-energy in the high temperature
limit. The first term on the right hand side of equation (15) can
lead to the following photon rate \cite{thoma} using the Boltzmann
distribution functions instead of a quantum mechanical ones:
\begin{equation}
\left(2E\frac{dn}{d^3pd^4x}\right)_1= 
\frac{5\alpha\alpha_s\lambda^2_q\lambda_g}{9\pi^2}T^2e^{-E/T}
\left[ln\left(\frac{4ET}{k^2_c}\right)-1.42\right].
\end{equation}
\noindent
Here $\alpha$ and $\alpha_s$ are the electromagnetic and the
strong interaction coupling constants. The second term in equation (15)
will give, under the Boltzmann approximation, the following contribution
to the photon rate:
\begin{widetext}
\begin{eqnarray}
\hspace{-1cm}
\left(2E\frac{dn}{d^3pd^4x}\right)_2 &=&\\ \nonumber
&& \frac{10\alpha\alpha_s}{9\pi^4}T^2e^{-E/T}
 \left\lbrace \lambda_q\lambda_g\left(1-\lambda_q\right)
 \left[1-2\gamma+2ln\left(4ET/k^2_c\right)\right]
+\lambda_q\lambda_q\left(1-\lambda_g\right)
 \left[-2-2\gamma+2ln\left(4ET/k^2_c\right)\right]\right\rbrace,
\end{eqnarray}
\end{widetext}

The total photon production rate $2E\frac{dn}{d^3pd^4x}$ can be obtained
by adding  equations (16-17), is required to be convoluted with the
space time evolution of the heavy-ion collision. We define \cite{thoma}

\begin{eqnarray}
\left(2\frac{dn}{d^2p_\bot dy}\right)_{y,p_\bot}&=&\int d^4x\left(2E\frac{dn}{d^3pd^4x}\right)\\
&=&Q\int_{\tau_0}^{\tau_1}
d\tau ~\tau \int_{-y_{nuc}}^{y_{nuc}}dy^{'}
\left(2E\frac{dn}{d^3pd^4x}\right)\nonumber
\end{eqnarray}

where the times after the maximum overlap of the nuclei are $\tau_0$ and
and $\tau_1$ and $y_{nuc}$ is the rapidity of the nuclei. $Q$ is the
transverse cross-section of the nuclei and $p_\bot$ is the photon momentum
in direction perpendicular to the collision axis.
For a $Au$ nucleus $Q \sim 180 fm^2$.
The quantity $\left(2E\frac{dn}{d^3pd^4x}\right)$ is  Lorentz invariant
and it is evaluated in the local rest frame in equation (18). 
The photon energy in this frame, i.e in the frame comoving with
the plasma, can be written as $p_\bot cosh(y-y^\prime)$.

\begin{figure}[h]
\begin{tabular}{cc}
\resizebox{40mm}{40mm}{\includegraphics{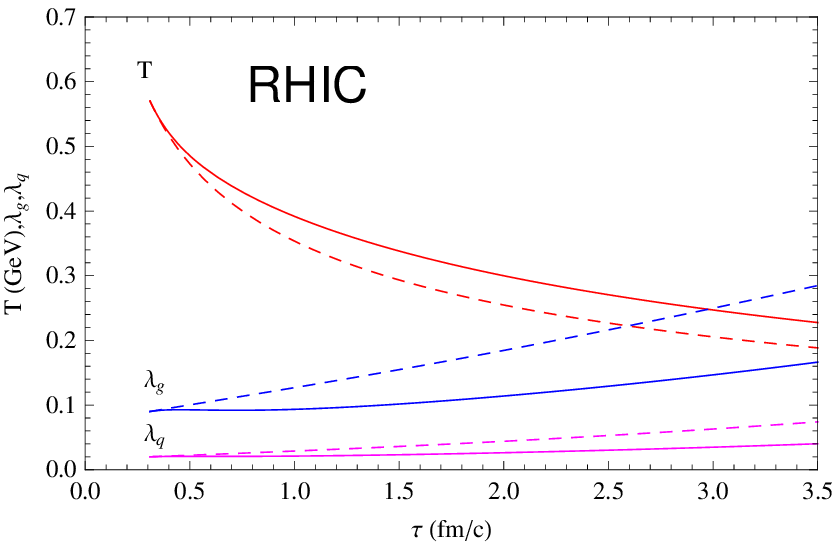}}&
\resizebox{40mm}{40mm}{\includegraphics{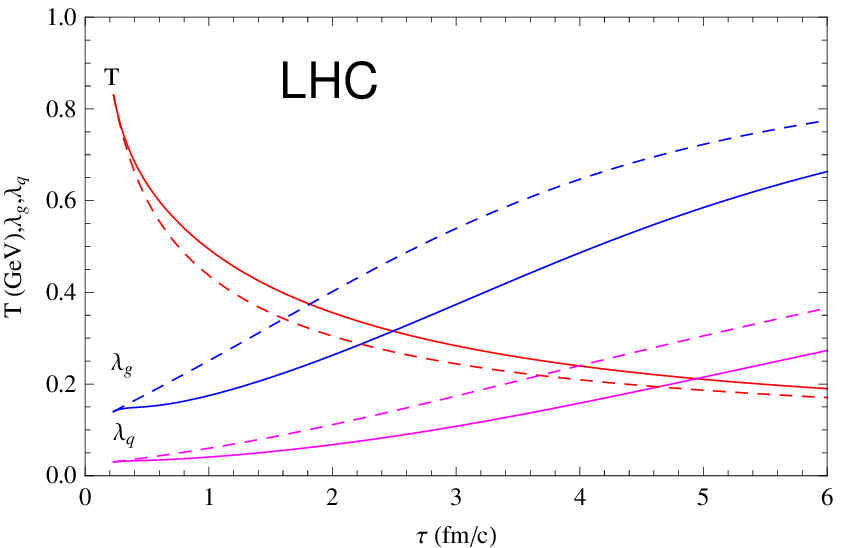}}
\end{tabular}
%\begin{center}
%\includegraphics[width=6cm,height=6cm]{mergedgraph.eps}
%\end{center}
\caption{Temperature, gluon fugacity and quark fugacity for RHIC
 and LHC.
Solid lines indicate the case with the shear viscosity, while
the dashed lines correspond to the case without viscosity}
\end{figure}

In Figure (1), we have shown $T$,$\lambda_g,\lambda_q$ as function of time.
We have solved the equations (9-12) together with the initial conditions at $\tau_{iso}$ from HIJING Monte Carlo model \cite{HIJING}. Which are $\lambda^{0}_{g}\,=\,0.09$, $\lambda^{0}_{q}\,=\,0.02$ and $T_{o}\,=\,0.57 GeV$ for RHIC with $\tau_{iso}\,=\,0.31 fm/c$ and $\lambda^{0}_{g}\,=\,0.14$, $\lambda^{0}_{q}\,=\,0.03$ and $T_{o}\,=\,0.83 GeV$ for LHC with $\tau_{iso}\,=\,0.23 fm/c$. Presence of the causal viscosity decreases the fall of temperature due to expansion
and the chemical nonequilibrium. However if one considers the first
order theory, there can be unphysical instability. Fugacity of gluons
and quarks increase more slowly due to the presence of the viscosity
compared to the cases when no viscous effects were included. This is
because the chemical equilibration is reached here with falling
of the temperature. 
The temperature can decrease due to the expansion and chemical
nonequilibration. The lowering of $T$ can help in attaining chemical
equilibrium and which in turn will increase the rate at which
the fugacities reach their equilibrium values. 
Inclusion of the viscosity will slowdown the falling rate of the temperature. Consequently the fugacities
will take more time to reach their equilibrium values.

\begin{figure}[h]
\begin{tabular}{cc}
\resizebox{40mm}{40mm}{\includegraphics{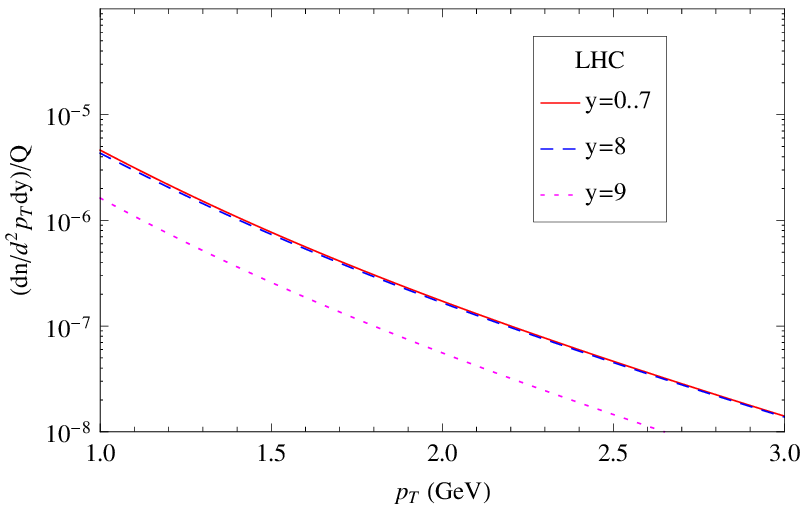}}&
\resizebox{40mm}{40mm}{\includegraphics{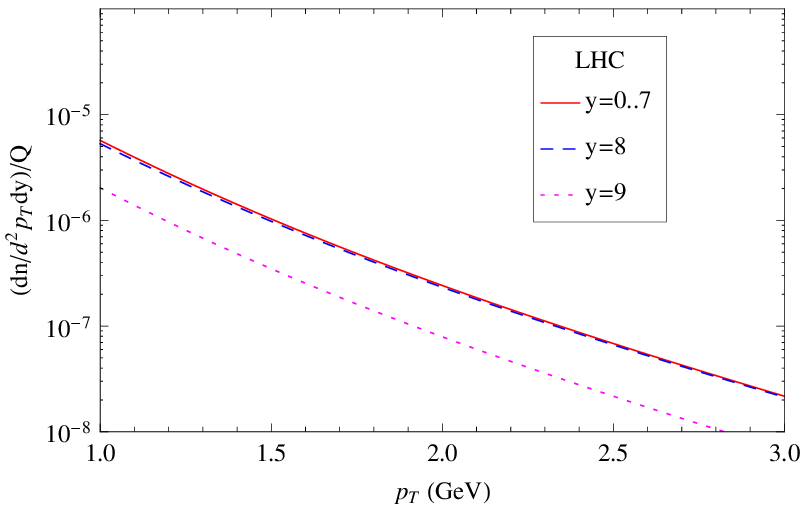}}
\end{tabular}
%\begin{center}
%\includegraphics[width=6cm,height=6cm]{LHCphoton.eps}
%\includegraphics[width=6cm,height=6cm]{LHC-photon-visc.eps}
%\end{center}
\caption{\em{(Left panel)Photon rate for different rapidities in LHC ($y_{nuc}\,=\,8.8$). (Right panel) Same with the inclusion of viscosity.}}
\end{figure}
We plot photon spectra  by using 
equation (\ref{taupimronga}) for $\tau_{\pi}$ in solving equations (9-12). The figures (2-3) compare the case without viscosity with
the case of finite shear viscosity.

Figure (2) shows the photon spectra emitted at fixed rapidities 
as a function of transverse momenta $p_\bot$. The photon flux
is normalized with the transverse size of the colliding
nuclei($Q$). For LHC we take: $\tau_0\,=\,0.5 fm/c$, $\tau_1\,=6.25 fm/c$ and $y_{nuc}\,=\,8.8$. We use equation (\ref{taupimronga}) for $\tau_{\pi}$ in solving equations (9-12). The figure compares the case without viscosity with
the case of finite shear viscosity \cite{emxg08}.

Figure (3) shows the comparison similar to that of figure (2) but with
a set of initial conditions for RHIC: 
$\tau_0\,=\,0.7 fm/c$,$\tau_1 \,=\, 4 fm/c$ and 
$y_{nuc}\,=\,6.0$. For $\alpha_s = 0.3$, shear viscosity to entropy density ratio $\eta/s\sim 0.29$.
Figures (2-3) show that viscous effects enhance the
photon flux by a factor (1.5-2).
\begin{figure}
\begin{tabular}{cc}
\resizebox{40mm}{40mm}{\includegraphics{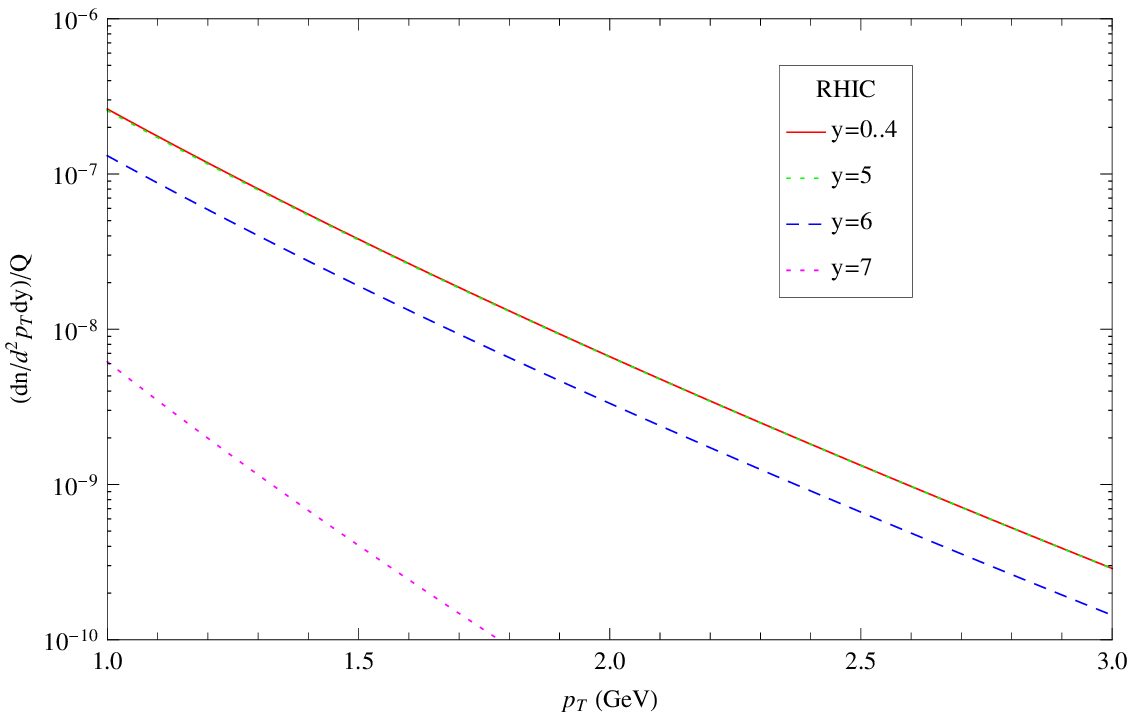}}&
\resizebox{40mm}{40mm}{\includegraphics{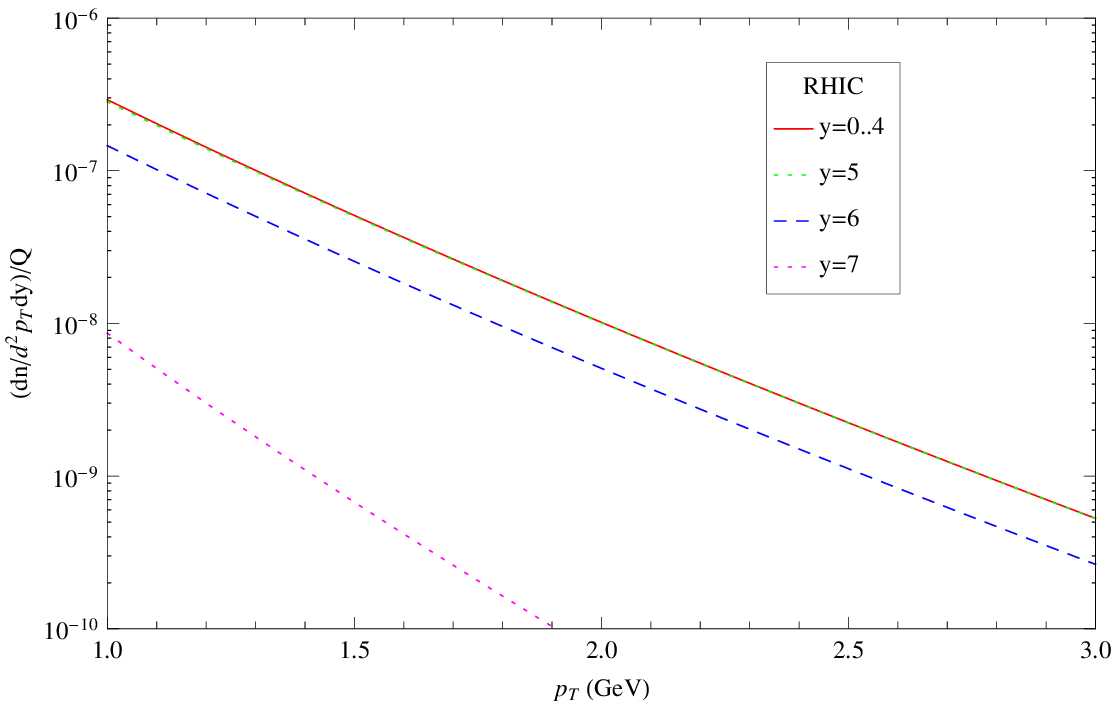}}
\end{tabular}
%\begin{center}
%\includegraphics[width=6cm,height=6cm]{RHICphoton.eps}
%\includegraphics[width=6cm,height=6cm]{RHIC-photon-visc.eps}
%\end{center}
\caption{\em{(Left panel)Photon rate for different rapidities in RHIC ($y_{nuc}$=6.0). (Right panel) Same with the inclusion of viscosity.}}
\end{figure}

Finally, we compare the photon fluxes calculated using  
equation (\ref{taupimronga}) with the fluxes calculated using
the kinetic viscosity ($\tau_{\pi}=3/{2\pi T}$). 
Figure (4) shows the photon flux calculated using the kinetic
viscosity prescription for LHC and RHIC.
However, we do not find any significant change in the flux 
for the results obtained using equation (\ref{taupimronga}).

\begin{figure}
\begin{tabular}{cc}
\resizebox{40mm}{40mm}{\includegraphics{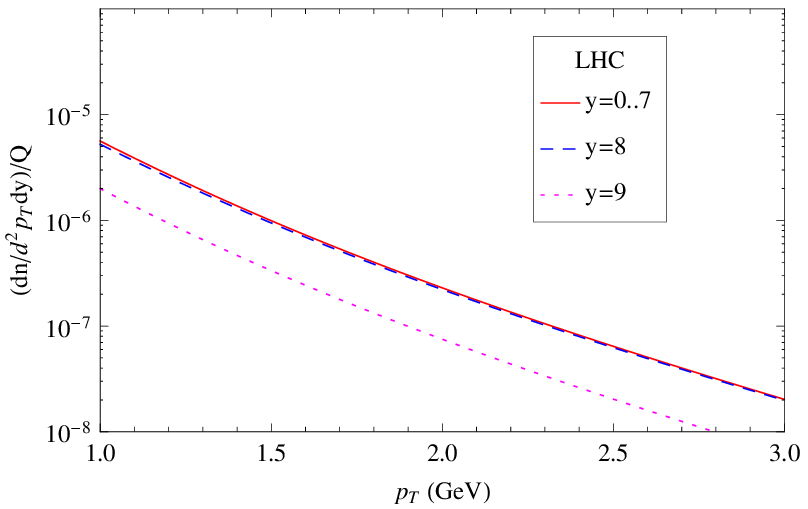}}&
\resizebox{40mm}{40mm}{\includegraphics{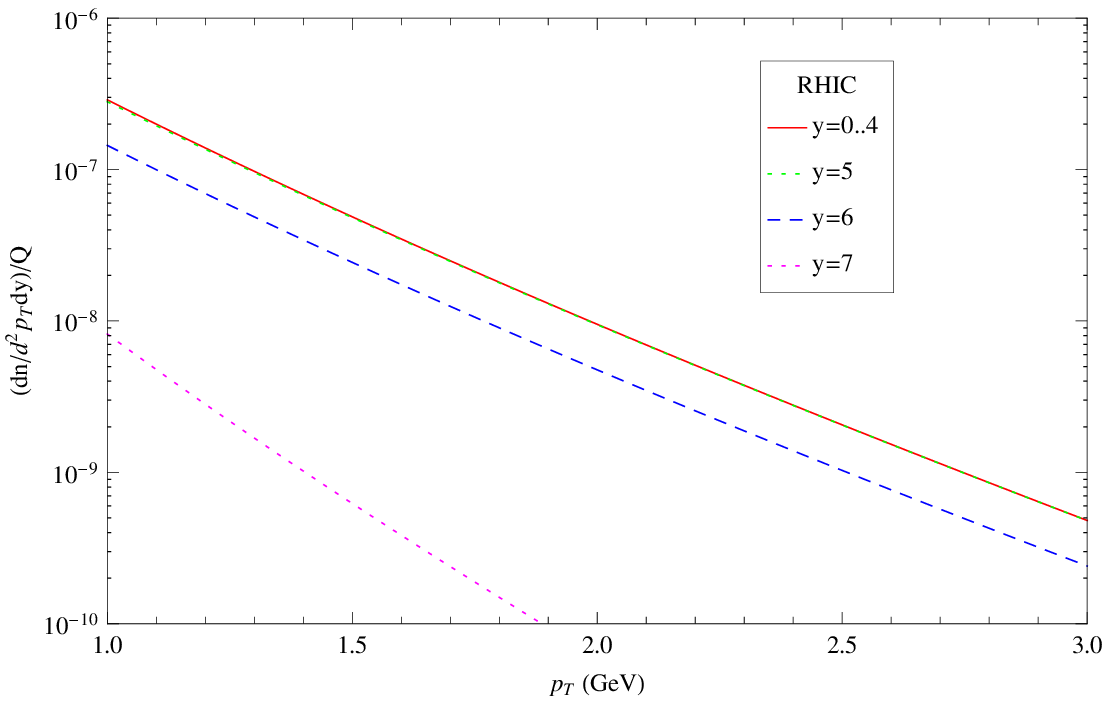}}
\end{tabular}
%\begin{center}
%\includegraphics[width=6cm,height=6cm]{LHC-photon-visc-kinetic.eps}
%\includegraphics[width=6cm,height=6.2cm]{RHIC-photon-visc-kinetic.eps}
%\end{center}
\caption{\em{Photon rate for different rapidities in RHIC ($y_{nuc}$=6.0) and LHC ($y_{nuc}$=8.8) with kinetic viscosity.}}
\end{figure}

In conclusions, we have studied the second order dissipative 
hydrodynamics with chemical nonequilibration. We
find that the effect of viscosity enhancing  
the photon flux by a factor ranging between
1.5-2 for the parameter space relevant for LHC and RHIC.
Our results are in a broad qualitative agreement with
the results obtained in Ref.\cite{chau00} using the
first order theory. We also find that
the two viscosity prescriptions with inelastic scattering
\cite{emxg08} and the one involving
elastic collisions only using kinetic theory give similar results
for the photon production rate.

\end{document}